\documentstyle[aps,pra,preprint]{revtex}
\begin{document}
\title{Preparation of multi-party entanglement of individual photons and atomic
ensembles }
\author{Guo-Ping Guo\thanks{%
Electronic address: harryguo@mail.ustc.edu.cn }, Guang-Can Guo}
\address{Key Laboratory of Quantum Information, University of Science and Technology\\
of China, Chinese Academy of Science, Hefei, Anhui, P. P. China, 230026}
\maketitle

\begin{abstract}
An experimental feasible scheme is proposed to generate
Greenberger-Horne-Zeilinger (GHZ) type of maximal entanglement.
Distinguishing from the previous schemes, this entanglement can be chosen
between either atomic ensembles (stationary qubit) or individual photons
(flying qubit), according to the difference applications we desire for it. 
The physical requirements of the scheme are moderate and well fit the
present experimental techinque.

PACS number(s): 03.65.Ud, 03.67.-a, 42.50.Gy, 42.50.-p
\end{abstract}

Entanglement of many parties is of fundamental interest to test quantum
mechanics against local hidden theory\cite{zh1,zh2}. Furthermore, it has
many practical applications in various quantum information processing tasks
such as quantum cryptography\cite{zh3}, computer\cite{zh4}, and teleportation%
\cite{zh5}. It is also believed that with more subsystems entangled, quantum
non-locality becomes more striking\cite{zh1,du2}, and quantum entanglement
is more useful in actual applications\cite{du3,du4,du5}. Great attention has
been directed to get more and more subsystems entangled. In theory, numerous
schemes have been proposed to generate multi-party entanglement with cavity%
\cite{guo,zh}, ion traps\cite{so}, spontaneous parametric down converter
(SPDC)\cite{du6,du9}, and indistinguishable atoms in Bose-Einstein
condensates\cite{du10}. In experiment, there are reports of demonstration of
four-photon entanglement in SPDC\cite{du9} and four-particle entanglement in
ion traps\cite{zh11}.

Based on the stimulated Raman emission of atomic ensemble and the associated
collective enhancement effect, atomic ensemble has been shown useful in long
distance quantum communication\cite{d,g}. And there are also protocols to
generate multi-party entanglement between atomic ensembles\cite{du} or to
entangle atomic ensembles with Stokes photon\cite{guo1}. The inherent robust
to realistic noise and imperfections of the collective states of the atomic
ensembles enable them as a preferable choose for the stationary qubit of
quantum computation and information. While the manipulation convenience with
linear optics for the photon polarization states makes individual photon a
suitable candidate for flying or transmitting qubit of the quantum
computation and information.

Here we propose a scheme to generate GHZ type of maximal entanglement.
Different from the previous multi-party entanglement preparation protocol
with atomic ensembles, the entangled subsystems of the present scheme can be
chosen as either atomic ensembles (stationary qubit) or individual photons
(flying qubit), according to the difference application. However, due to the
collective enhancement effect of the atomic ensemble Raman procession, the
physical requirements of the present scheme are still moderate and well fit
the current experimental technique.

As most works with atomic ensembles\cite{d,g,du,guo1}, the basic element of
our scheme is also an ensemble of many identical atoms, whose experimental
realization can be either a room-temperature atomic gas or a sample of cold
trapped atoms. The relevant structure of the atom is shown in Fig. 1. From
the three levels $\left| g\right\rangle ,\left| r\right\rangle ,\left|
l\right\rangle ,$ we can define two collective atomic operators $s=(1/\sqrt{%
N^a})\sum\limits_{i=1}^{N^a}\left| g\right\rangle _i\left\langle s\right| $
with $s=r,l,$ where $N^a\gg 1$ is the total atom number. Initially the atoms
are optically pumped to the ground state $\left| g\right\rangle ,$which is
effectively a vacuum state $\left| 0\right\rangle $ of the operators $r,l.$
A basis of the polarization qubits can be defined from the states $\left|
R\right\rangle =r^{\dagger }\left| 0\right\rangle $ and $\left|
L\right\rangle =l^{\dagger }\left| 0\right\rangle $, which both have an
experimentally demonstrated long coherence time\cite{du12,du14,du15,du16}.
In Raman processing, the atomic ensemble is transferred from the ground
state $\left| g\right\rangle $ to the excited state $\left| e\right\rangle $
by a classical laser (the pump light) with Rabi frequency $\Omega .$
Shortly, this excited state will transit to the two metastable states $%
\left| r\right\rangle $ and $\left| l\right\rangle $ with equal
probabilities. In these transitions $\left| e\right\rangle \rightarrow
\left| r\right\rangle $ and $\left| e\right\rangle \rightarrow \left|
l\right\rangle $, the atomic ensemble will respectively emit a Stokes
photon, which is horizontally or vertically polarized. Due to the collective
enhanced coherent interaction, these excitation modes $r$ and $l$ can be
respectively transferred to optical excitation modes $h$ and $v$ with high
precision. Then they can be detected by single-photon detectors, even for a
free-space ensemble, which has been demonstrated in both in theory\cite{du17}
and in experiments\cite{du14,du15}.

As pointed out in the paper\cite{guo1}, EPR entangled state $\left| \Psi
^1\right\rangle =(r_a^{\dagger }h_p^{\dagger }+l_a^{\dagger }v_p^{\dagger })/%
\sqrt{2}\left| 0_{ap}\right\rangle $ between atomic ensemble and Stokes
photon can be prepared with a short off-resonant laser pulse in this atomic
ensemble system. This preparation succeeds with a small probability $p$ for
each Raman drive pulse, which can be controlled by adjusting light-atom
interaction time $t_\Delta $ and pulse length\cite{d,du}. Generally, the
atomic ensemble and the Stokes photons can be totally written in the state

\begin{equation}
\left| \Psi \right\rangle _1=(I+p^{\frac 12}H+\sum_{j=2}^\infty \frac{(p^{%
\frac 12}H)^j}{j!})\left| 0_{ap}\right\rangle _1,
\end{equation}
where $I$ is the identity operator, $H=$ $(r_a^{\dagger }h_p^{\dagger
}+l_a^{\dagger }v_p^{\dagger })/\sqrt{2}$, and $\left| 0_{ap}\right\rangle $
is the vacuum state of the whole system. Here $h^{\dagger }$($v^{\dagger }$)
represents horizontal (vertical) mode creation operator of Stokes photon.
For convenience we leave this state unnormalized. It is obvious that the
probability to produce $m$ Stokes photons from an atomic ensemble decays
exponentially with the number $m$. The preparation for this state has
inherent resistance to noise and is well based on the current technology of
laser manipulation\cite{d,du}. Many applications, such as individual photons
quantum memory, can be expected from this novel entanglement between atomic
collective state (stationary qubit) and the individual photon polarization
state (flying qubit)\cite{g,guo1}.

In order to generate multi-party entanglement, we can first prepare $n$ pair
of entanglement between atomic ensembles and Stokes photons. A simply way is
to illuminate $n$ atomic ensembles in turn with a pump classical light. Thus
the whole system of the atomic ensembles and Stokes photons can be prepared
in the state (which is un-normalized) 
\begin{equation}
\left| \Phi \right\rangle ^{ap}=\prod_{i=1}^n\left| \Psi \right\rangle
_i=\prod_{i=1}^n(I+p^{\frac 12}H+\sum_{j=2}^\infty \frac{(p^{\frac 12}H)^j}{%
j!})\left| 0_{ap}\right\rangle _i,
\end{equation}
where the subscript $i$ represents the $i$th atomic ensemble. In the
expanding of this state, the terms involving $n$ photons all have a
coefficient of $p^{\frac n2}$. With this state, we can thereby generate
multi-particle entanglement of either $n$ atomic ensembles or $n$ photons.

In the following preparing procedure, we will employ a multi-photon GHZ
states analyzer with linear optics as shown in Fig. 3. In this optical
setup, the input photon in the mode $h_{p_i}^{\dagger }$ (or $%
v_{p_i}^{\dagger }$) can be transferred into mode $(h_{D_h^i}^{\dagger }+$ $%
v_{D_v^i}^{\dagger })/\sqrt{2}$ (or $(h_{D_h^{i-1}}^{\dagger }-$ $%
v_{D_v^{i-1}}^{\dagger })/\sqrt{2}$) , where we have assumed the notation $%
0\equiv n$ for the detector's subscript. Assume that there is only one
photon in each input. When there are coincidence clicks between $n$
detectors of this $n$-photon GHZ analyzer, and there are even (or odd)
number of $D_V^i$ among these $n$ click detectors, the $n$ photons are
obviously measured in the state $\left| M\right\rangle ^{+}$(or $\left|
M\right\rangle ^{-}$), where $\left| M\right\rangle ^{\pm }=(1/\sqrt{2}%
)(\prod_{i=1}^nh_{p_i}^{\dagger }\pm \prod_{i=1}^nv_{p_i}^{\dagger })\left|
0\right\rangle ^p$. Here $\left| 0\right\rangle ^p$ represents the total
vacuum state of the $n$ photons. Generally, this linear optics setup can
distinguish the states $\left| M\right\rangle ^{\pm }$ from the other states
of the $n$-party GHZ states. When $n=2$, it becomes a Bell states analyzer
which can divide the four Bell states into three classes: $\left| \Phi
^{+}\right\rangle =\frac 1{\sqrt{2}}(h_{p_1}^{\dagger }h_{p_2}^{\dagger
}+v_{p_1}^{\dagger }v_{p_2}^{\dagger })\left| 0\right\rangle ^p$, $\left|
\Phi ^{-}\right\rangle =\frac 1{\sqrt{2}}(h_{p_1}^{\dagger }h_{p_2}^{\dagger
}-v_{p_1}^{\dagger }v_{p_2}^{\dagger })\left| 0\right\rangle ^p$ and $\left|
\Psi ^{\pm }\right\rangle =\frac 1{\sqrt{2}}(h_{p_1}^{\dagger
}v_{p_2}^{\dagger }\pm v_{p_1}^{\dagger }h_{p_2}^{\dagger })\left|
0\right\rangle ^p$. Thereby this Bell states analyzer with linear optics can
be straightforwardly applied in quantum information processing such as
quantum teleportation. For the multi-photon case, much more applications can
be expected for the $n$-photon GHZ states analyzer\cite{p,gg}. In the
following, we will show how to generate multi-atomic-ensemble or
multi-photon entanglement, with the above $n$-photon GHZ states analyzer.

As the $n$ atomic ensembles and their emitting Stokes photons are totally in
the state $\left| \Phi \right\rangle ^{ap}$, we respectively input those
Stokes photons from the $n$ atomic ensembles into the above $n$-photon GHZ
states analyzer as Fig. 3. When there are coincidence clicks between $n$
detectors of this $n$-photon GHZ analyzer and there are even (or odd) number
of $D_V^i$ among these $n$ click detectors, the input Stokes photons are
measured in the states$\left| M\right\rangle ^{\pm }$ or $\left| \;M^{\prime
}\right\rangle $, where $\left| \;M^{\prime }\right\rangle $ represents the
states that there are more than one photon in some optics input but still
has the same coincidence clicks as the case that there is one and only one
Stokes photon in each input. With this measurement, the residual $n$ atomic
ensembles of the state $\left| \Phi \right\rangle ^{ap}$ is projected into
the states 
\begin{equation}
\rho ^a=\left| \Phi \right\rangle _{\pm }^a\left\langle \Phi \right| +\rho
_{vac}^a.
\end{equation}
Here $\left| \Phi \right\rangle _{\pm }^a=$ $(1/\sqrt{2})(%
\prod_{i=1}^nr_{a_i}^{\dagger }\pm \prod_{i=1}^nl_{a_i}^{\dagger })\left|
0\right\rangle ^a$ is $n$-atomic-ensemble GHZ type maximal entangled state,
which results from the case that the input Stokes photons are measured in
the states$\left| M\right\rangle ^{\pm }$. And $\rho _{vac}^a$ represents
the $n$ atomic ensemble states when the Stokes photons are projected in the
state $\left| \;M^{\prime }\right\rangle $. This is the case that some
ensembles have more than one excitation and some others have no excitation.
It is easy to see that the effect of the detector inefficiencies and loss of
excitation can be also combined into the term $\rho _{vac}^a$. Note that for
any practical application of the multi-party entanglement, the state
preparation should be succeeded by a measurement of the polarization of the
excitation on each ensemble\cite{zh2,du2,du3,du4}. There we only keep these
results, for which excitation appears on each ensemble. Then the state $\rho
^a$ can effectively yield $n$-atomic-ensemble GHZ type maximal entanglement
as the states $\left| \Phi \right\rangle _{\pm }^a$ in any application.

As we can respectively transfer the atomic ensembles excitation modes $r$
and $l$ to optical excitation modes $h$ and $v$, the entanglement between $n 
$ atomic ensembles can be directly transferred to $n$ photons. Then we can
get $n$-photon GHZ entanglement state straightly from the state $\rho ^a$.
Alternatively, we can measure the $n$ atomic ensembles of the state $\left|
\Phi \right\rangle ^{ap}$in the basis $\left| N\right\rangle ^{\pm
}=(\prod_{i=1}^nr_{a_i}^{\dagger }\pm \prod_{i=1}^nl_{a_i}^{\dagger })/\sqrt{%
2}\left| 0\right\rangle ^a$ to project the corresponding Stoke photons into
GHZ entanglement state. This measurement can be done when the excitations of
the $n$ atomic ensembles of the state $\left| \Phi \right\rangle ^{ap}$ is
transferred to $n$ optical excitations, and then measured with the same GHZ\
state analyzer. Similarly, those optical modes transferred from the atomic
ensembles excitation modes can be measured in the state $\left|
M\right\rangle ^{\pm }$ with post-selection. This equals to measure the $n$
atomic ensembles in the basis $\left| N\right\rangle ^{\pm }$. Thereby the
remaining $n$ Stokes photons of the state $\left| \Phi \right\rangle ^{ap}$
are projected in an entangled state, which can effectively yield $n$-photon
GHZ type maximal entanglement as the state $\left| \Psi \right\rangle ^p=(1/%
\sqrt{2})(\prod_{i=1}^nh_{p_i}^{\dagger }\pm \prod_{i=1}^nv_{p_i}^{\dagger
})\left| 0\right\rangle ^p=\langle N^{\pm }\left| \Phi \right\rangle ^{ap}$
in any application.

It has been shown that the inherent resilience to noise of the collective
states of atomic ensemble can enable it as a well qualified candidate for
stationary and register qubits of quantum information and computation\cite
{d,g}. On the other hand, the light is an ideal carrier of quantum
information, and the individual photon polarization modes can be
conveniently manipulated. According to the different applications, the
present scheme can elegantly prepare either $n$-atomic-ensemble or $n$%
-photon entanglement in one experimental setup. Although post-selection is
still needed as in most of multi-party entanglement generation schemes such
as SPDC scheme\cite{du6,du9}, these states can yield effectively GHZ
entanglement whenever they are put into applications.

We now give a brief discussion on the efficiency and the practical
implementation of this proposal. As shown in the paper\cite{d,du}, we can
control the probability of getting a Stokes photon $p$ $\sim 10^{-2}$ for a
Raman driving pulse with a short light-atom interaction time $t_\Delta $.
Then the $n$ atomic ensembles has a probability of order of $p^n$ to produce 
$n$ Stokes photons. When we only post-select the case that every ensemble
has one excitation, the probability to get $n$-party entanglement is $%
p^n/2^{n-1}$ which equals $10^{-6}$ in the case $n=3$. Thus with a typical
repetition frequency $f_p=10^7Hz$ for the Raman pulses, we can get $10$
pairs of three-party GHZ entangled states per second. The ensembles should
be prepared into ground state before each round of Raman processing. It is
also noted that the present entangling scheme suffers from the fast
exponential degradation of the efficiency as most protocols with
post-selection.

As the unknown phase differences between the Stokes photons are fixed by the
optics setup, we can effortlessly balance them with some phase plates\cite
{d,du}. In the practical implementation, we can safely neglect the dark
counts of the single-photon detectors in the coincidence detections\cite{du6}%
. The transferring of the atomic ensemble excitation mode to the optical
mode can be high efficient. As long as the preparing procedure is
accomplished in a time no longer than the coherence times of the atomic
collective states $T_{pre}\preceq $ $ms$, we can also safely neglect the
noise of the non-stationary phase drift induced by the pumping lase or by
the optical channel.

In conclusion, we have proposed an experimental feasible scheme to prepare
multi-party GHZ type entanglement between either atomic ensembles
(stationary qubit) or individual photons (flying qubit) with post-selection
in one experiment setup. Due to the collective enhancement effect of the
atomic ensemble Raman procession, the physical requirements of this scheme
are moderate and well fit for the current experimental technique.

This work was funded by National Fundamental Research Program(2001CB309300),
National Natural Science Foundation of China, the Innovation funds from
Chinese Academy of Sciences, and also by the outstanding Ph. D thesis award
and the CAS's talented scientist award entitled to Luming Duan.

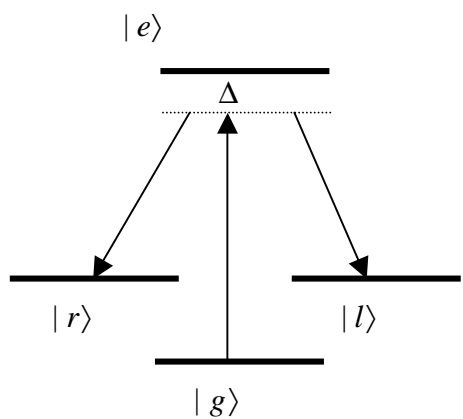
Figure 1: The relevant level structure with $\left| g\right\rangle $ the
ground state, $\left| e\right\rangle $ the excited state, and $\left|
r\right\rangle $ , $\left| l\right\rangle $ the two metastable state for
storing a qubit. The transition $\left| g\right\rangle \rightarrow \left|
e\right\rangle $ is coupled by a classical laser (the pump light) with Rabi
frequency $\Omega ,$ followed with two equal probability transitions $\left|
e\right\rangle \rightarrow \left| r\right\rangle $ and $\left|
e\right\rangle \rightarrow \left| l\right\rangle ,$ where right-handed and
left-handed rotation forward-scattered Stokes photons are emitted
respectively. For convenience, we assume off-resonant coupling with a large
detuning $\Delta .$

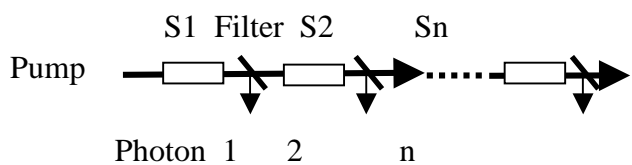
Figure 2: Schematic drawing of the experimental setup for the generation of
the entanglement state$\left| \Phi \right\rangle ^{ap}$. The
frequency-selective filter separates the pump light from the Stokes photon.
The ensembles are prepared in the ground state before each round of Raman
procession.

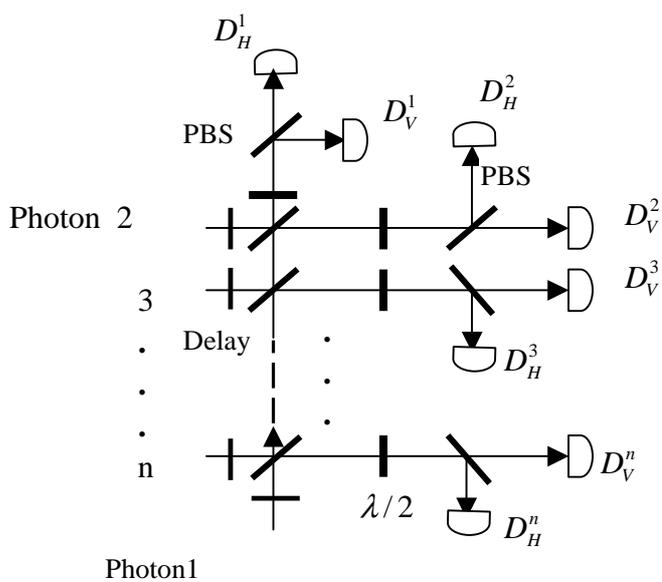
Figure 3: Schematic of the experimental setup for the measurement $\left|
M\right\rangle ^{\pm }=(1/\sqrt{2})(\prod_{i=1}^nh_{p_i}^{\dagger }\pm
\prod_{i=1}^nv_{p_i}^{\dagger })\left| 0\right\rangle ^p$of the $n$ Stokes
photon in the state$\left| \Phi \right\rangle ^{ap}$. The polarizing beam
splitters (PBS) reflect vertical photons and transmit horizontal photons. We
can adjust the arrival time of the $n$ photons with the delay plates. The $%
\lambda /2$ plates are employed to rotate the polarization of the Stokes
photon $i$ through $45^0$ to transfer the photon $h^{\dagger }$ mode into $%
(h^{\dagger }+v^{\dagger })/\sqrt{2}$ and $v^{\dagger }$ mode into $%
(h^{\dagger }-v^{\dagger })/\sqrt{2},$ where $\lambda $ is the wavelength of
those photons. This optical setup transfers the mode $h_{p_i}^{\dagger }$
into $(h_{D_h^i}^{\dagger }+$ $v_{D_v^i}^{\dagger })/\sqrt{2}$ and the mode $%
v_{p_i}^{\dagger }$into $(h_{D_h^{i-1}}^{\dagger }-$ $v_{D_v^{i-1}}^{\dagger
})/\sqrt{2}$, where we have assumed the notation $0\equiv n$ for the
detector's subscript. Then for the case that $n$ photon are detected and
even (odd) number of detectors $D_v^i$ click, the $n$ photon are projected
into state $\left| M\right\rangle ^{+}$ (or $\left| M\right\rangle ^{-}$).


\begin{references}
\bibitem{zh1}  J. S. Bell, Physics (Long Island City, N.Y.) 1. 195 (1965).

\bibitem{zh2}  D. M. Greenberger, M. A. Horne, A. Shimony and A. Zeilinger,
Am. J. Phys. 58, 1131 (1990).

\bibitem{zh3}  A. K. Ekert, Phys. Rev. Lett. 67, 661 (1991).

\bibitem{zh4}  D. Deutsch and R. Jozsa, Proc. R. Soc. London A 439, 553
(1992).

\bibitem{zh5}  C. H. Bennett {\it et al}., Phys. Rev. Lett. 70, 1895 (1993).

\bibitem{du2}  N. D. Mernin, Phys. Rev. Lett. 65, 1838 (1990).

\bibitem{du3}  J. Bollinger, W. M. Itano, D. Wineland and D. Heinzen, Phys.
Rev. A 54, 4649 (1996).

\bibitem{du4}  D. Gottesman and I. L. Chuang, Nature 402, 390 (1999).

\bibitem{du5}  M. A. Nielsen and I. L. Chuang, {\it Quantum Computation and
Quantum Information, }(Cambridge University Press, UK, 2000).

\bibitem{guo}  G. P. Guo,{\it \ et. al. , }Phys. Rev. A 65, 042102 (2002).

\bibitem{zh}  S. B. Zheng, Phys. Rev. Lett. 87, 230404 (2002).

\bibitem{so}  C. A. Sackett {\it et al.}, Nature 404, 256 (2000).

\bibitem{du6}  D. Bouwmeester {\it et al.}, Phys. Rev. Lett. 82, 1345 (1999).

\bibitem{du9}  J. W. Pan {\it et al.}, Phys. Rev. Lett. 86, 4435 (2001).

\bibitem{du10}  A. Sorensen, L. M. Duan, J. I. Cirac and P. Zoller, Nature
406, 63 (2001).

\bibitem{zh11}  C. A. Sackett {\it et al.}, Nature 404, 256 (2000).

\bibitem{d}  L. M. Duan, M. D. Lukin, J. I. Cirac, P. Zoller, Nature 414,
413 (2001).

\bibitem{g}  G. P. Guo and G. C. Guo, Phys. Lett. A 318, 337 (2003).

\bibitem{du}  L. M. Duan, Phys. Rev. Lett. 88, 170402 (2002).

\bibitem{guo1}  G. P. Guo and G. C. Guo, {\it Quantum Information and
Computation} (QIC) Vol. 3 No. 6 627-634 (2003).

\bibitem{p}  J.-W. Pan and A. Zeilinger, Phys. Rev. A 57, 2208 (1998). 

\bibitem{gg}  G. C. Guo and G. P. Guo, Phys. Rev. A 68, 044303 (2003).

\bibitem{du12}  B. Julsgaard, A. Kozhekin and E. S. Polzik, Nature 431, 400
(2001).

\bibitem{du14}  D. F. Phillips {\it et al.}, Phys. Rev. Lett. 86, 783 (2001).

\bibitem{du15}  C. Liu, Z. Dutton, C. H. Behroozi and L. V. Hau, Nature 409,
490 (2001).

\bibitem{du16}  J. F. Roch {\it et al.}, Phys. Rev. Lett. 78, 634 (1997).

\bibitem{du17}  M. Fleischhauer and M. D. Lukin, Phys. Rev. Lett. 84, 5094
(2000).

{\bf Figure Captions:}
\end{references}
\end{document}